\documentclass[aps,prd,12pt,notitlepage,
nofootinbib,superscriptaddress,showkeys,showpacs]{revtex4-1}

\usepackage{amstext,amsmath,amssymb,amsfonts,bbm,euscript,color,dsfont}
\usepackage{graphicx}

\usepackage[utf8]{inputenc}

\usepackage[english]{babel}
\usepackage[T1]{fontenc}
\usepackage[caption=false]{subfig}
\usepackage{epstopdf}
\usepackage{amsmath}
\usepackage{amsfonts}
\usepackage{amssymb}
\usepackage{latexsym}
\usepackage{hyperref}
\usepackage{comment}
\usepackage{xcolor}
\usepackage{bm}
\usepackage{bbold}
\usepackage{eufrak}
\usepackage{graphicx}
\usepackage{float} % for using [H]
\usepackage{multirow}
\usepackage{array}
\usepackage{booktabs}
\usepackage{float}
\usepackage{verbatim}
\usepackage{tikz}
\usepackage[normalem]{ulem}
\usepackage{xcolor}

\begin{document}

\title{Kinematic and Dynamic Parameters of Anisotropic Models with Inflaton, Without \textit{A Priori} Relation Between the Scale Factors and with \textit{A Posteriori} Isotropization.}

\author{M. Perovano}
\email{maressaperovanof@gmail.com}
\affiliation{Universidade Federal do Esp\'irito Santo, Vit\'oria, ES, Brazil}

\author{W. Spalenza}
\email{wspalenza@gmail.com}
\affiliation{Instituto Federal do Esp\'{i}rito Santo, IFES - Cariacica, ES, Brazil}

\author{B. B. da Silva}
\email{brenobsilva@gmail.com}
\affiliation{Universidade Federal do Esp\'irito Santo, Vit\'oria, ES, Brazil}

\author{H. Belich}
\email{belichjr@gmail.com}
\affiliation{Universidade Federal do Esp\'irito Santo, Vit\'oria, ES, Brazil}

\author{S.V.B. Gon\c{c}alves}
\email{sergio.v.goncalves@ufes.br}
\affiliation{Universidade Federal do Esp\'irito Santo, Vit\'oria, ES, Brazil}

\begin{abstract}
In this work, we present a detailed review of the calculations concerning homogeneous and anisotropic cosmological models, with particular emphasis on the Kantowski-Sachs cosmological model. As an innovative contribution, we derive an exact solution for the field equations, expressed in terms of a general form for the material content of the universe. We obtain the expressions for the energy density in this general case, as well as investigate the specific scenario involving the presence of a scalar field. We examine the evolutionary dynamics of the model that encompasses Hubble, density, and deceleration parameters. Additionally, we calculate the kinematic parameters associated with the model, namely: the scalar expansion $\Theta$, the shear scalar $\sigma$, the rotation parameter $\omega$, and the parameter $R_{\alpha\beta}U^{\alpha}U^{\beta}$, also known as the Raychaudhuri scalar or convergence condition, all of these parameters expressed within the framework of the Raychaudhuri equation. Instead of employing a pre-established relation between the scale factors of the model, we allow them to evolve freely, without \textit{a priori} conditions. Thus, we can verify that the isotropic nature of the studied scenario manifests itself rapidly, leading to the traditional isotropic model.

\end{abstract}  

\maketitle

%%%%%%%%%%%%%%%%%%%%%%%%%%%%%%%%%%%%
\section{Introduction}
%%%%%%%%%%%%%%%%%%%%%%%%%%%%%%%%%%%%

The understanding of the spatiotemporal evolution of the cosmos on its largest scales stands as one of the central pillars of contemporary cosmology. Although the $\Lambda$CDM paradigm — grounded in the cosmological principle that postulates the homogeneity and isotropy of the universe on large scales — has achieved remarkable success in interpreting a vast array of observational data \cite{Kolb, Weinb, Misner, Peeb, caroll}, it still harbors ambiguities and empirical inconsistencies that motivate the investigation of more comprehensive theoretical frameworks. In this scenario, homogeneous but anisotropic cosmological models assume a relevant role, as they allow testing the stability of deductions based on strictly isotropic geometries, in addition to acting as a theoretical platform for examining processes such as the isotropization process and the evolution of large-scale cosmological structures \cite{Ehlers, Wu}.

Anisotropic geometries, which admit different expansion rates along different spatial axes, have been the focus of deep investigations in recent decades. In this regard, the Bianchi classification offers a structured formalism for characterising homogeneous spatiotemporal manifolds \cite{ellis, bairagi}. Among the Bianchi scenarios, those of class I stand out, associated with spatially flat universes, as well as the Kantowski-Sachs (KS) metric \cite{KS01, KS02}, which exhibits a spatial topology of the form $\mathbb{R} \times S^2$ and is defined by two independent scale factors. This model, in particular, has attracted attention due to its phenomenology in general, encompassing everything from the ability to simulate the interior region of black holes \cite{con, sime, barb} to providing a favorable environment for the analysis of quantum effects in the domain of cosmology \cite{qc04}.

Several recent studies have focused on evaluating the KS model from a dynamical systems perspective and the qualitative profile of its trajectories in phase space. In \cite{coley01}, for example, spatially homogeneous and closed cosmological scenarios composed of a perfect fluid coupled to a scalar field with an exponential potential are investigated. The solutions obtained demonstrate a multiplicity of dynamics driven by the choice of parameters, ranging from evolutions that fall into a final singularity to trajectories in continuous expansion or contraction that tend toward isotropy and exhibit inflationary phases. In a similar perspective, \cite{coley02} examines Einstein-\ae ther geometries with spherical symmetry that evolve toward spatially homogeneous KS models, using appropriate variables, identifying states that include inflationary attractors and sources characterized by zero curvature and maximum shear. On the other hand, \cite{chaubey} performs a qualitative study of Bianchi cosmologies of types I, III, and KS in the presence of a scalar field and ordinary matter, showing that, within the KS framework, the trajectories are mostly unstable — except in specific parametric configurations — and that the accelerated expansion regime detected in the recent era may correspond to certain fixed points of the considered dynamical system. Furthermore, \cite{hoogen} explores a set of anisotropic cosmological models in the context of Einstein-\ae ther gravity through the dynamical systems formalism, verifying that the equilibrium points of the KS model refer to anisotropic vacuum solutions associated with a positive deceleration parameter. Continuing this line, \cite{mohandas} addresses the non-trivial coupling between the shear of the aether field and the scalar field, mapping the possible future asymptotic behaviors, whose stability proves to be dependent on the magnitude of the shear coupling constant.

In addition to the mentioned contributions, the specialized literature presents a wide range of investigations based on anisotropic cosmologies \cite{shep, Andre, Vajk, KS03, leon, ger, Chakraborty}, extending from the Bianchi approaches \cite{ellis, bairagi} to the KS formalism itself. The Kantowski-Sachs metric, specifically, is characterized by two parameters $a(t)$ and $b(t)$, which represent the scale factors governing the temporal dynamics of the universe in the radial and transverse directions, respectively. The existence of two autonomous scale factors inherently introduces an anisotropic dynamics. In the pioneering study by Kantowski and Sachs \cite{KS01}, the geometry was originally treated under the premise of a pressureless fluid ($p=0$). Preceding this work, Kompaneets and Chernov \cite{KC} had already analyzed an anisotropic cosmological scenario consisting of dust and thermal radiation. In his doctoral thesis, Kantowski expanded these results to contemplate scenarios with a cosmological constant, photon gas ($p = \rho/3$), and stiff matter ($p = \rho$), that is, barotropic equations of state in the form $p = w \rho$, with $w = 0, 1, 1/3$ \cite{KS02}.

The Kantowski-Sachs spacetime possesses singular properties. Its solutions in the classical regime are well documented in multiple contexts, as are its quantum formulations \cite{con, sime, barb}. Additionally, the KS line element can be equivalent to the Schwarzschild vacuum solution in the region interior to the event horizon, making it a convenient model for describing the interior of black holes and examining their quantum states and entropy. Models with the KS structure are also valuable in studying the behavior of supplementary degrees of freedom in quantum models of the universe. As an example, quantization in the minisuperspace of Bianchi I brings to light divergences of interpretation between the Bohm-de Broglie and many-worlds formulations of quantum mechanics, where the quantum solution manifests non-unitarity and temporal evolution in the norm of the quantum state \cite{qc04}.

In this work, we propose to examine the Kantowski-Sachs metric in order to solve the Einstein field equations considering a generic matter distribution for the cosmos, rather than restricting our approach to particular fluids and their equations of state a priori. In this way, we broaden and systematize the approaches proposed in the aforementioned articles, in which different material contents were addressed in isolation, without a unified framework. Through this more comprehensive formulation of the field equations in the KS metric, it becomes feasible to simultaneously accommodate various material components, such as stiff matter, relativistic radiation, dust, scalar fields, among others. Additionally, we determine the kinematic cosmological parameters associated with the geometry, including the Hubble rate and the temporal profile of the two scale factors. A relevant differential of the present methodology lies in the investigation of both scale factors without imposing prior hypotheses that establish a functional dependence between them. Our calculations treat both scale factors as independent temporal functions, enabling a more consistent verification of their evolutions. Finally, we analyze the integrability of the Raychaudhuri equation and derive the scalar expansion analytically, examining the geodesic congruence, interpreted in the present context as the family of integral curves of a vector field defined on a four-dimensional Lorentzian manifold.

The article is structured as follows. In Section 2, we present our results for the Kantowski-Sachs model and for the Bianchi type I and III models with local rotational symmetry (LRS). The kinematic parameters are studied in Section 3. We develop the expansion dynamics of the model, including the Hubble, density, and deceleration parameters, in Section 4. The free expansion of the scale factors, $a(t)$ and $b(t)$, and their behaviors are presented in Section 5. Our conclusions and future perspectives are outlined in Section 6. Our conclusions and future perspectives are outlined in Section 6. Due to the complexity of the ordinary differential equations involved, numerical methods were required to obtain the solutions. Therefore, the differential equations were solved numerically using the fourth-fifth-order Runge–Kutta–Fehlberg method (RKF45) \cite{press2007numerical}, allowing for the determination of the solutions and the subsequent generation of the plots. Subsequently, the numerical results were subjected to curve fitting using the least-squares method \cite{draper1998}. The quality of the fits was evaluated using the statistical metric coefficient of determination, $R^2$, with fits yielding values closer to 1, considered more suitable \cite{Montgomery2021}.

%%%%%%%%%%%%%%%%%%%%%%%%%%%%%%%%%%%%%%%%%%%%%%%%%%%%%%%%%%%%%%%%%%%%%%%%%%%%%%%%%%%%%
\section{General Solutions for the Models}
%%%%%%%%%%%%%%%%%%%%%%%%%%%%%%%%%%%%%%%%%%%%%%%%%%%%%%%%%%%%%%%%%%%%%%%%%%%%%%%%%%%%%%%

In this section, our goal is to present the anisotropic expressions related to two types of geometry: for the case of a closed universe ($\kappa=+1$), we have the Kantowski-Sachs model; for a flat universe ($\kappa=0$), we have the Bianchi type I model with local rotational symmetry (LRS). Additionally, we will consider the case of an open universe ($\kappa=-1$), corresponding to the Bianchi type III LRS model.

\subsection{Kantowski-Sachs Model}

Our starting point is the work of Kantowski and Sachs \cite{KS01}, who proposed a metric characterized by two distinct scale factors: one acting in the radial direction and another in the angular direction. This metric is given by
\begin{equation} \label{KS00b}
ds^2 = dt^2 - a(t)^2 dr^2 - b(t)^2 \left(d\theta^2 + \sin^2 {\theta} \,\, d\varphi^2 \right)\quad,
\end{equation}
where \( a(t) \) and \( b(t) \) are functions of cosmic time that describe the evolution of the universe. The presence of two scale factors with potentially independent evolutions naturally introduces a scenario of anisotropy. To process the model, we insert the metric into the Einstein field equations, which, in natural units ($c = \hbar = 8\pi G = 1$), are written as
\begin{equation} \label{Einstein}
R_{\mu \nu} - \frac{1}{2} R g_{\mu \nu} = T_{\mu \nu}\quad.
\end{equation}
The energy-momentum tensor is composed of the sum of the contributions from the perfect fluid and the scalar field, $T_{\mu \nu} = T^{f}_{\mu \nu} + T^{\phi}_{\mu \nu}$, with
\begin{eqnarray}
T^{f}_{\mu \nu} &=& (\rho + p) U_\mu U_\nu - p g_{\mu \nu}, \\
T^{\phi}_{\mu \nu} &=& \phi_{,\mu} \phi_{,\nu} - g_{\mu \nu} \left[\frac{1}{2}\phi_{,\kappa} \phi^{,\kappa} - V(\phi)\right]\quad.
\end{eqnarray}
Thus, we obtain the following field equations for the Kantowski-Sachs metric
\begin{eqnarray}
2 \frac{\dot{a} \dot{b}}{a b} + \left( \frac{\dot{b}}{b} \right)^2 + \frac{1}{b^2} &=& \rho + \rho_{\phi}\quad, \label{alpharho1} \\
2\frac{\ddot{b}}{b} + \left(\frac{\dot{b}}{b} \right)^2 + \frac{1}{b^2} &=& -p - p_{\phi}\quad, \label{alpharho2} \\
\frac{\ddot{a}}{a} + \frac{\ddot{b}}{b} + \frac{\dot{a} \dot{b}}{a b} &=& -p - p_{\phi}\quad. \label{alpharho3}
\end{eqnarray}
Here, the pressure and energy density of the scalar field are defined by
\begin{eqnarray}
\rho_{\phi} &=& \frac{1}{2} {\dot{\phi}}^2 + V(\phi), \\
p_{\phi} &=& \frac{1}{2} {\dot{\phi}}^2 - V(\phi)\quad,
\end{eqnarray}
with the scalar field potential given by
\begin{equation}
V(\phi) = \frac{1}{2} m^2 \phi^2 - \frac{\lambda}{4!} \phi^4\quad.
\end{equation}
Note that in expressions \eqref{alpharho1} and \eqref{alpharho2} the term $\frac{1}{b^2}$ appears, which is responsible for the influence of positive spatial curvature, characterizing the closed geometry of the Kantowski-Sachs model.

\subsection{Bianchi Type I LRS Model}

The Bianchi type I LRS metric also presents two scale factors, $a(t)$ and $b(t)$, now associated with Cartesian coordinates, and is defined by
\begin{equation} \label{LRS-B1}
ds^2 = dt^2 - a(t)^2 dx^2 - b(t)^2 (dy^2 + dz^2)\quad.
\end{equation}
This model constitutes an anisotropic but homogeneous cosmological solution of the Einstein equations, describing a universe with rotational symmetry around a preferred axis, while maintaining spatial homogeneity. It belongs to the Bianchi classification of homogeneous cosmological models, categorized by the structure of their spatial symmetry groups. The model presents an axis of rotational symmetry (generally along the ${z}$ direction), so that the physical properties are invariant under rotations around this axis. The two transverse directions (${x}$ and ${y}$) are equivalent, yet distinct from the longitudinal direction (${z}$). The difference between $a(t)$ and $b(t)$ introduces an anisotropy in the cosmic expansion, allowing the expansion to be faster in one direction than in the others.

The field equations associated with this metric take the form
\begin{eqnarray}
2\frac{\dot{a} \dot{b}}{a b} + \frac{\dot{b}^2}{b^2} &=& \rho + \rho_{\phi}\quad, \label{LRSBI1} \\
2\frac{\ddot{b}}{b} + \frac{\dot{b}^2}{b^2} &=& -p - p_{\phi}\quad, \label{LRSBI2}\\
\frac{\ddot{a} }{a} + \frac{\ddot{b}}{b} + \frac{\dot{a} \dot{b}}{a b} &=& -p - p_{\phi}\quad. \label{LRSBI3}
\end{eqnarray}
The difference between these expressions and the field equations of the Kantowski-Sachs model, given by \eqref{alpharho1}, \eqref{alpharho2}, and \eqref{alpharho3}, lies solely in the absence of the curvature term $\frac{1}{b^2}$. It is precisely this absence that confers upon the Bianchi type I LRS model a spatially flat geometry.

\subsection{Bianchi Type III LRS Model}

The Bianchi type III LRS model is an anisotropic metric that possesses two scale factors \cite{Fadragas, Karagiorgos}, being given by
\begin{equation} \label{KS00c}
ds^2 = dt^2 - a(t)^2 dr^2 - b(t)^2 \left(d\theta^2 + \sinh^2 {\theta} \,\, d\varphi^2 \right)\quad.
\end{equation}
This line element arises as a case of open geometry in the article by Kantowski-Sachs \cite{KS01}. Applying the Einstein equations, expression \eqref{Einstein}, to this metric, we obtain the following field equations
\begin{eqnarray}
2\frac{\dot{a} \dot{b}}{a b} + \left( \frac{\dot{b}}{b} \right)^2 - \frac{1}{b^2} &=& \rho + \rho_{\phi}\quad, \label{LRSBIII1} \\
2\frac{\ddot{b}}{b} + \left(\frac{\dot{b}}{b} \right)^2 - \frac{1}{b^2} &=& -p - p_{\phi}\quad, \label{LRSBIII2} \\
\frac{\ddot{a}}{a} + \frac{\ddot{b}}{b} + \frac{\dot{a} \dot{b}}{a b} &=& -p - p_{\phi}\quad. \label{LRSBIII3}
\end{eqnarray}
The difference between the above equations and those of Kantowski-Sachs, expressions \eqref{alpharho1} and \eqref{alpharho2}, is that the open geometry manifests through the sign change in the term related to curvature, $-\frac{1}{b^2}$.

\subsection*{A Preliminary Analysis}

Unlike other studies that establish \textit{a priori} a relation between the two scale factors of the model, here we allow the scale factors to evolve freely. With this, we obtain a more realistic behavior of the process involving the initial anisotropic model and its subsequent isotropization. Here, $a(t)$ and $b(t)$ are obtained from a system of coupled nonlinear differential equations. For the KS model, for example, we have
\begin{eqnarray}
\ddot\phi &=& - 3H\dot\phi - m^2\phi + \lambda\phi^3\quad, \nonumber\\
\ddot a &=& a\left[\frac{\dot b^2}{2b^2} - \frac{\dot a~\dot b}{a~b} - \frac{1}{2}p_t + \frac{1}{2b^2}\right]\quad, \nonumber\\
\ddot b &=& b\left[- \frac{1}{2}p_t - \frac{\dot b^2}{2b^2} - \frac{1}{2b^2}\right]\quad, \nonumber\\
\dot\rho &=& -3H(p + \rho)\quad, \label{ks_nonlinear_1}
\end{eqnarray}
where the total pressure and energy density are $p_t = p_{\phi} + p$ and $\rho_t = \rho_{\phi} + \rho$, respectively, with $p_{\phi} = \frac{1}{2}\dot\phi^2 - V(\phi)$, $\rho_{\phi} = \frac{1}{2}\dot\phi^2 + V(\phi)$, and the fluid equation of state is $p = w \rho$, where $w = 0, \, 1, \, 1/3$. The scalar field potential is $V(\phi) = \frac{1}{2}m^2\phi^2 - \frac{1}{4}\lambda\phi^4$ and $H = \frac{1}{3}\left(\frac{\dot a}{a} + 2\frac{\dot b}{b}\right)$ is the Hubble parameter.

\begin{figure}[H]
\centering
    \begin{minipage}{0.32\textwidth}
        \centering
        \includegraphics[width=\linewidth]{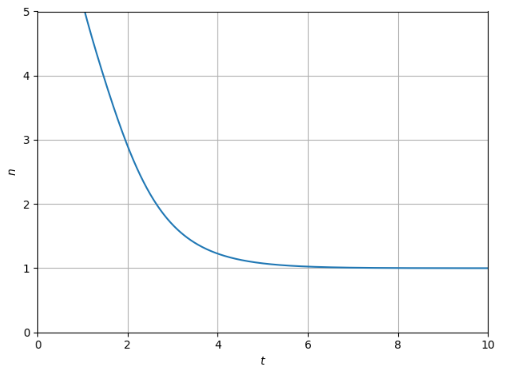}
        \textbf{(A)}
        \label{fig1}
    \end{minipage}
    \begin{minipage}{0.32\textwidth}
        \centering
        \includegraphics[width=\linewidth]{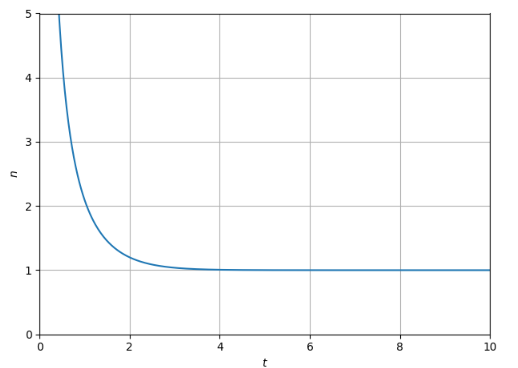}
        \textbf{(B)}
        \label{fig2}
    \end{minipage}
    \begin{minipage}{0.32\textwidth}
        \centering
        \includegraphics[width=\linewidth]{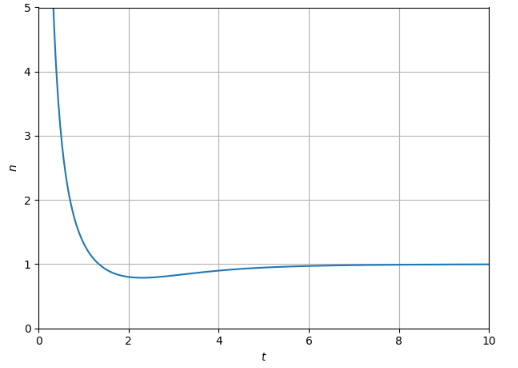}
        \textbf{(C)}
        \label{fig3}
    \end{minipage}
    \caption{Numerical comparison for the power $n(t)$ obtained from the KS (A), Bianchi I (B), and Bianchi III (C) models. The temporal variation of $n$ evidences that the \textit{ansatz} $a=b^n$, with constant $n$, to describe the complete solutions of the dynamical system, can be worked in a more general manner.}
    \label{figura_1}
\end{figure}

Some references use an \textit{ansatz} to fix solutions to cosmological problems in the KS metric (see, for example, \cite{fator}). One such \textit{ansatz} has the form
\begin{equation}
a = b^n\quad, \label{ansatz_abn}
\end{equation}
generally employed when inflation is not involved in the model in question, being an attempt to give a form to the nonlinear solutions. From there, they substitute this \textit{ansatz} into the nonlinear equations (\ref{ks_nonlinear_1}) and analyze values for $n$, where, in general, the resulting functions exhibit polynomial or exponential character. Although this approach seems functional at first, there may be a certain degree of arbitrariness in it, since the genuine solutions of the system are not necessarily polynomial. Furthermore, no constant term is introduced in the equation that could determine a linear coefficient. A pertinent question is: is the exponent $n$ constant, that is, must it remain constant for all values of time $t$? We can analyze this with a simple calculation. We take the derivative of $a(t)$ with respect to $b(t)$ from Eq. (\ref{ansatz_abn}) and then isolate $n$, using the equation itself, obtaining $n = \frac{b}{a}\frac{\dot{a}}{\dot{b}}$. If the relation between $a(t)$ and $b(t)$ is a straight line, this value is constant; otherwise, it is not. With this, using the results of $a(t)$ and $b(t)$ obtained from solving the coupled ODEs, Eq. (\ref{ks_nonlinear_1}), we observe that $n$ is not constant, as illustrated in Figure \ref{figura_1}. We also extended this analysis to the other two metrics addressed here in this study. 
This indicates that the \textit{ansatz} $a=b^n$ is a tentative assumption that could be further refined, as a more complex functional form may be required given that the solution arises from a system of coupled nonlinear equations. Several studies have also adopted alternative types of \textit{ansatz}, as discussed in \cite{gonccalves2022kantowski}, \cite{Kantowski1966},\cite{xanthopoulos1992kantowski} among others.

%%%%%%%%%%%%%%%%%%%%%%%%%%%%%%%%%%%%%%%%%%%%%%%%%%%%%%%%%%%%%%%%%%%%%%%%
\section{Analysis of the Scale Factor Fitting from the Nonlinear Model}
\label{FE}
%%%%%%%%%%%%%%%%%%%%%%%%%%%%%%%%%%%%%%%%%%%%%%%%%%%%%%%%%%%%%%%%%%%%%%%%

The solutions of the system of four nonlinear ODEs, in arbitrary units, provide data that do not correspond to simple analytical functions. To overcome this impasse, we can employ these data, which describe curves, to perform a fitting treatment, here restricted to the scale parameters $a(t)$ and $b(t)$, but which can be extended to other parameters of the theory. For this task, we use a nonlinear fitting method, which solves curve fitting problems employing the nonlinear least squares method. This method is widely used when the model depends nonlinearly on its parameters. The method can make use of an implementation of the Levenberg-Marquardt algorithm, which combines the gradient descent technique, effective far from the minimum, with the Gauss-Newton method, which is effective near the minimum it would be interesting to put a reference here at this point in the article \cite{burden2016numerical}.

The central idea of this work is to find a way to express the numerical solutions resulting from solving the system of ordinary differential equations obtained, and to try analytical combinations through curve fitting, where we analyze various combination attempts, as detailed below. Then, we employ the least squares method (LSM) in some variations described below.

To analyze the data from the ODE solutions, we developed a code that solved the coupled ODEs via initial value problem (IVP) using the RKF45 method with the fundamental variables: $t, \, a, \, b, \, \dot{a}, \, \dot{b},\, \rho, \, \dot{\rho}, \, \phi, \, \dot{\phi}$. From these data, we created a code that would read the data, perform the fitting of the original curve (from the ODE solutions), testing various types of functions in a database of polynomial and transcendental functions, and thus choose and statistically test the best function, or combination among at least three functions from this database. We used the statistical metric parameters: adjusted coefficient of determination $R^2$ and root mean square error (RMSE) \cite{devore2016probability}, which ranked the functions and combinations (from the database). The analysis was performed for three metric models, as mentioned earlier, and also for four types of fluids, to verify if the curve fitting would be adequate for any model. Therefore, these results allowed us to select the best function, which we present below.

%%%%%%%%%%%%%%%%%%%%%%%%%%%%%%%%%%%%%%%%%%%%%%%%%%%%%%%%%%%%%%%%%%
\subsection*{Case (A): Kantowski-Sachs for $ w = \frac{1}{3}$}
%%%%%%%%%%%%%%%%%%%%%%%%%%%%%%%%%%%%%%%%%%%%%%%%%%%%%%%%%%%%%%%%%%

\begin{figure}[!htbp]
    \centering
    \begin{minipage}{0.95\textwidth}
        \centering
        \includegraphics[width=\linewidth]{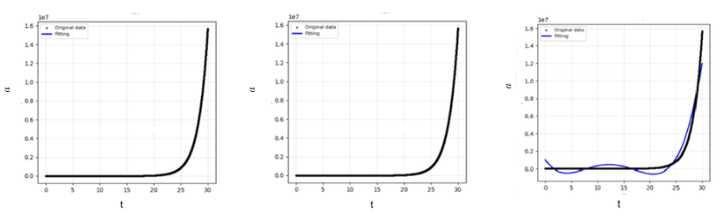}
        \caption{Comparison between three curves obtained via fitting: original data (in black) and fit (blue), whose best equation is $a(t) = 0.47102\,\exp(0.57736 t) - 1.07901$, with statistical metric of $R^2 = 0.999998$.}
           \label{figura_2}
    \end{minipage}
 
\end{figure}

\begin{figure}[!htbp]
    \centering
    \begin{minipage}{0.95\textwidth}
        \centering
        \includegraphics[width=\linewidth]{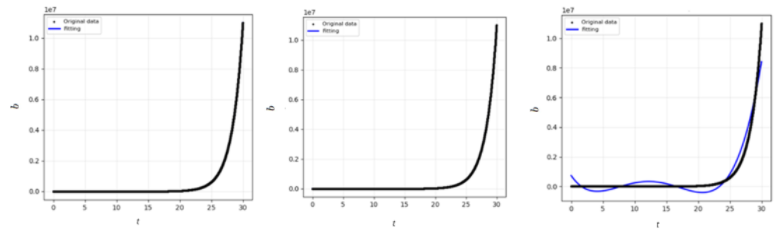}
        \caption{Comparison between three curves obtained via fitting: original data (in black) and fit (blue), whose best equation is $b(t) = 0.33329\,\exp(0.57736\,t) - 0.69682$, with statistical metric of $R^2 = 0.999998$.}
           \label{figura_3}
    \end{minipage}
 
\end{figure}

\begin{figure}[!htbp]
    \centering
    \begin{minipage}{0.45\textwidth}
        \centering
        \includegraphics[width=\linewidth]{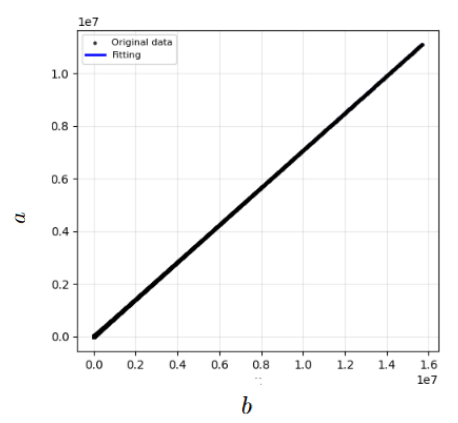}
    \end{minipage}
    \caption{Graph of the best curve, among three obtained via fitting, for the original data (in black) and the fit (blue). The best equation is $a(b) = 1.8618\times10^{-28}b^{4} - 5.8800\times10^{-21}b^{3} + 5.9699\times10^{-14}b^{2} + 0.6987b + 0.1655$, with statistical metric of $R^2 = 0.999999$.}
    \label{figura_4}
\end{figure}

\begin{figure}[!htbp]
    \centering
    \begin{minipage}{1.00\textwidth}
        \centering
        \includegraphics[width=\linewidth]{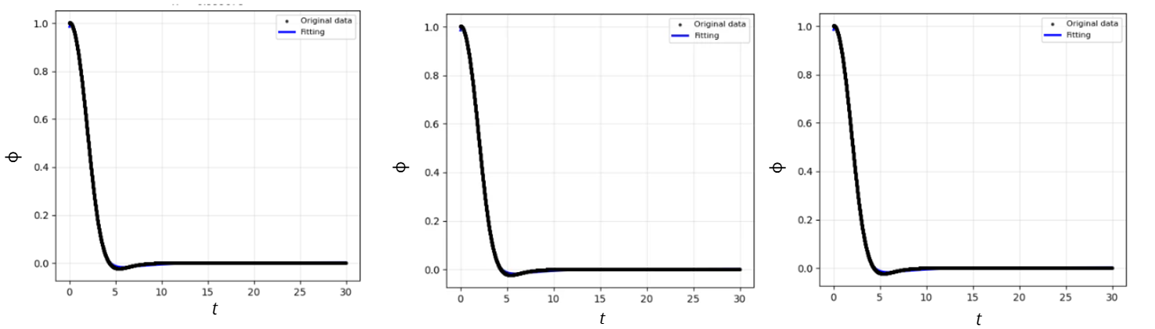}
        \caption{Fitting function with the form $\phi(t) =  1.8155t^3 + 2.3312t^2 + 1.8282t - 0.0303$, with statistical metric of $R^2 = 0.99998$.}
           \label{figura_5}
    \end{minipage}
 
\end{figure}

\begin{figure}[!htbp]
    \centering
    \begin{minipage}{1.00\textwidth}
        \centering
        \includegraphics[width=\linewidth]{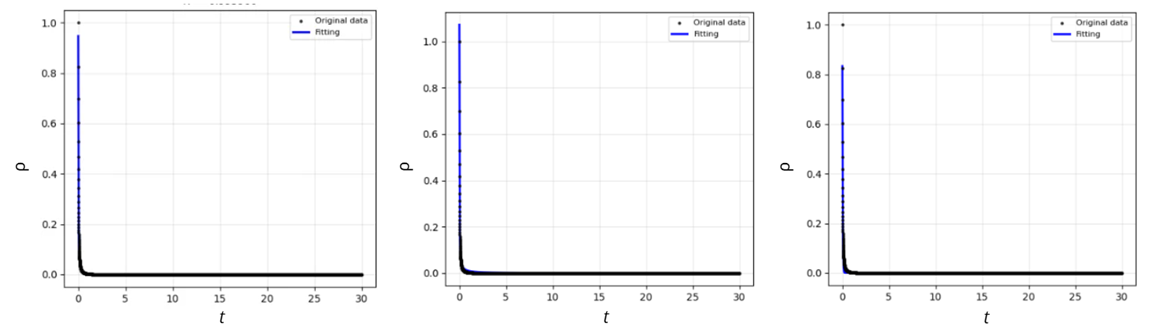}
        \caption{Fitting function with the form $\rho(t) =   3777.1395 - 3777.1391\,\tanh(15.1537t + 4.5561)$, with statistical metric of $R^2 = 0.99998$.}
          \label{figura_6}
    \end{minipage}
  
\end{figure}

For the Kantowski-Sachs cosmological model, we chose the thermodynamic fluid with the equation of state $ w=1/3$. In this case, we obtained the following best responses for $a \times t$ in Figure (\ref{figura_2}). For the scale parameter $b \times t$, the fit can be seen in Figure (\ref{figura_3}) , and in the figure description we indicate the best function. To confront the \textit{ansatz} used in some references, in equation (\ref{ansatz_abn}) which describes a $b \times a$ Figure \ref{figura_4}, we present the fit of this analysis. The fit for the field $\phi$ as a function of time $t$ provided slightly different results for each metric, with values quite close. Thus, the best function $\phi \times t$ is in Figure (\ref{figura_5}). Finally, we present the value of the fitted function for the density $\rho$ as a function of time $t$, obtained from the matter conservation equation, shown in Figure (\ref{figura_6}).



%%%%%%%%%%%%%%%%%%%%%%%%%%%%%%%%%%%%%%%%%%%%
\subsection*{Case (B): Bianchi Type I LRS}
%%%%%%%%%%%%%%%%%%%%%%%%%%%%%%%%%%%%%%%%%%%%

\begin{figure}[!htbp]
    \centering
    \begin{minipage}{\textwidth}
        \centering
        \includegraphics[width=\linewidth]{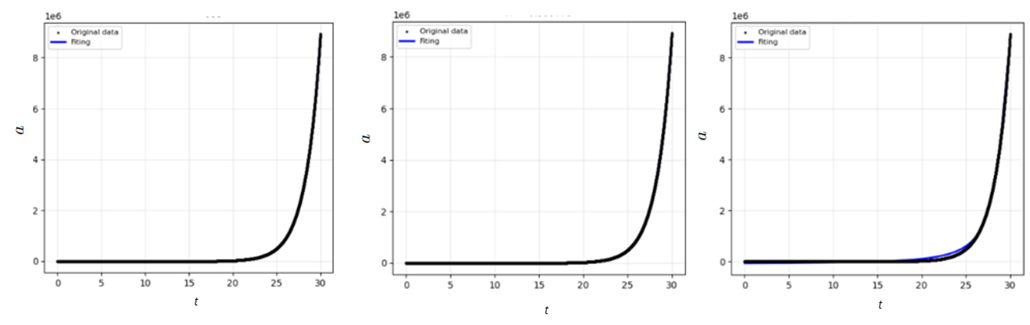}
        \caption{Comparison between three curves obtained via fitting: original data (in black) and fit (blue), whose best equation is $a(t) = 0.2686\,\exp(0.5774\,t) - 0.7769$, with statistical metric of $R^2 = 1.00000$.}
         \label{figura_7}
    \end{minipage}
   \end{figure}

\begin{figure}[!htbp]
    \centering
    \begin{minipage}{\textwidth}
        \centering
        \includegraphics[width=\linewidth]{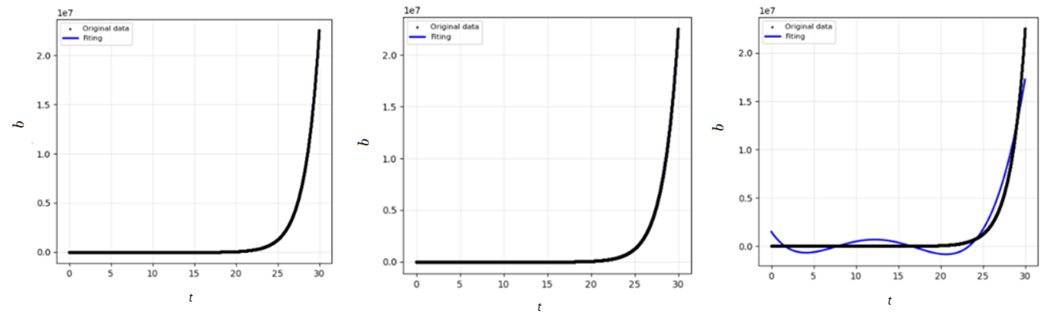}
        \caption{Comparison between three curves obtained via fitting: original data (in black) and fit (blue), whose best equation is $b(t) = 0.6778\,\exp(0.5774\,t) - 1.8972$, with statistical metric of $R^2 = 1.00000$.}
         \label{figura_8}
    \end{minipage}
   \end{figure}

\begin{figure}[!htbp]
    \centering
    \begin{minipage}{0.45\textwidth}
        \centering
        \includegraphics[width=\linewidth]{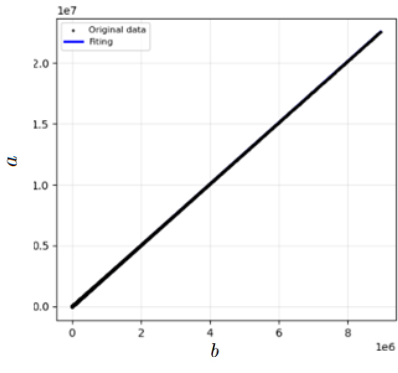}
    \end{minipage}
    \caption{Best curve, among three obtained via fitting, for the original data (in black) and the fit (blue). The best equation is $a(b) = 1.9014\times10^{-28}b^{4} - 5.9671\times10^{-21}b^{3} + 6.0186\times10^{-14}b^{2} + 0.70759b + 0.16450$, with statistical metric of $R^2 = 0.999998$.}
    \label{figura_9}
\end{figure}

\begin{figure}[!htbp]
    \centering
    \begin{minipage}{\textwidth}
        \centering
        \includegraphics[width=\linewidth]{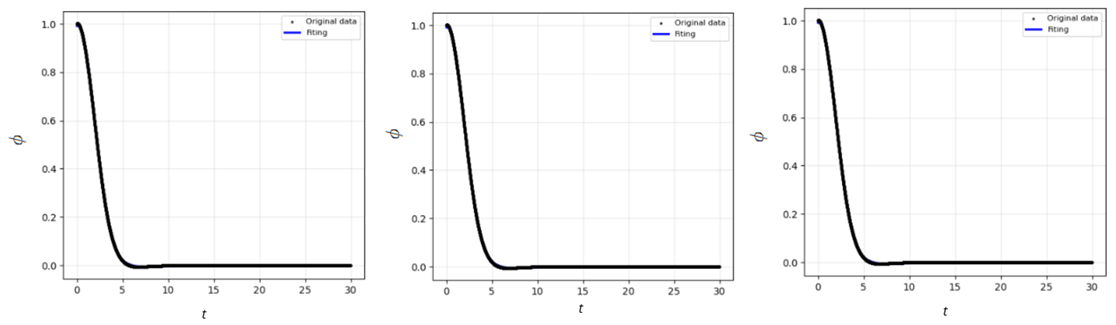}
        \caption{Fit for the scalar field: $\phi(t) = -0.0010 + 0.9967 \exp\left[-0.1524\left(t - 0.0871\right)^2\right] - 0.0010$, with statistical metric of $R^2 = 0.999998$.}
         \label{figura_10}
    \end{minipage}
   
\end{figure}

\begin{figure}[!htbp]
    \centering
    \begin{minipage}{\textwidth}
        \centering
        \includegraphics[width=\linewidth]{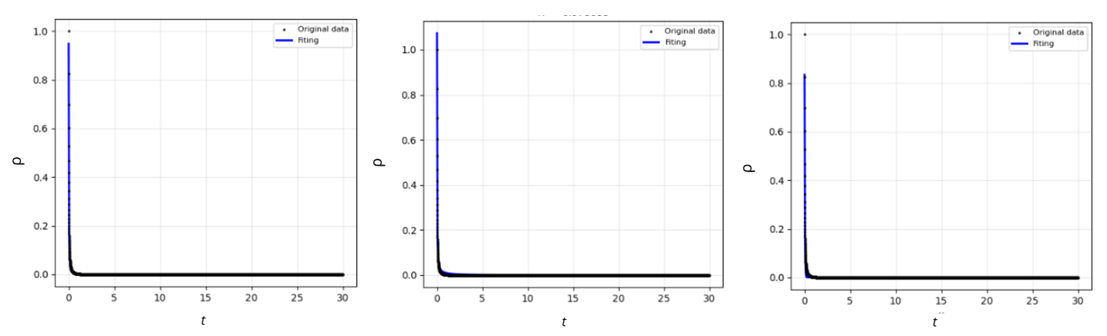}
        \caption{Fit for the density: $\rho(t) =  4348.7578 - 4348.7574\tanh(15.1842 t + 4.6262)$, with statistical metric of $R^2 = 0.96115$.}
         \label{figura_11}
    \end{minipage}
   
\end{figure}

We performed the same statistical treatment for the Bianchi type I LRS cosmological model. However, the fitted curve did not result in the same behavior as the KS model, which was expected, since the models describe distinct situations in the context of spacetime curvature. Here, we will present only the case of a fluid with $w = 1/3$, but we will mention the other cases later. The analysis for the graphs and curves is presented in Figures (\ref{figura_7}) and (\ref{figura_8}), for $a \times t$ and $b \times t$, respectively. Again, we confront the \textit{ansatz} used in some references, equation (\ref{ansatz_abn}), which describes the relation $b \times a$, whose fit is presented in the Figure \ref{figura_9}. For the scalar fields and density, (\ref{figura_10}) and (\ref{figura_11}), we have the graphs $\phi \times t$ and $\rho \times t$, respectively.
   

%%%%%%%%%%%%%%%%%%%%%%%%%%%%%%%%%%%%%%%%%%%%%
\subsection*{Case (C): Bianchi Type III LRS}
%%%%%%%%%%%%%%%%%%%%%%%%%%%%%%%%%%%%%%%%%%%%%

\begin{figure}[!htbp]
    \centering
    \begin{minipage}{\textwidth}
        \centering
        \includegraphics[width=\linewidth]{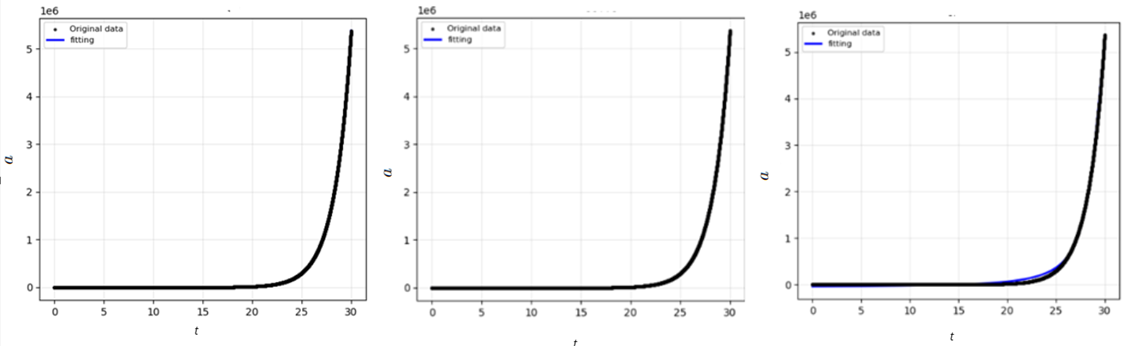}
        \caption{Comparison between three curves obtained via fitting: original data (in black) and fit (blue), whose best equation is $a(t) = 0.1614\,\exp(0.5774 t) - 0.7514$, with statistical metric of $R^2 = 1.00000$.}
        \label{figura_12}
    \end{minipage}
    
\end{figure}

\begin{figure}[!htbp]
    \centering
    \begin{minipage}{\textwidth}
        \centering
        \includegraphics[width=\linewidth]{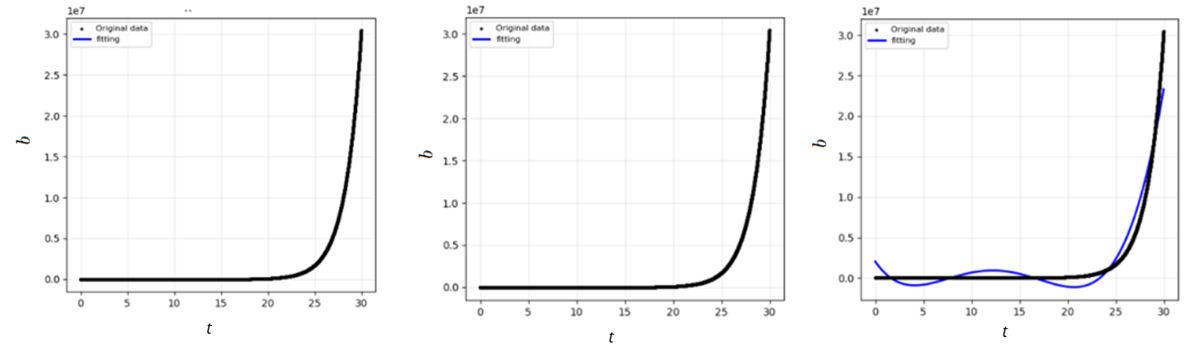}
        \caption{Comparison between three curves obtained via fitting: original data (in black) and fit (blue), whose best equation is $b(t) = 0.9157\,\exp(0.5774t)-5.0058$, with statistical metric of $R^2 = 1.00000$.}
          \label{figura_13}
    \end{minipage}
  
\end{figure}

\begin{figure}[!htbp]
    \centering
    \begin{minipage}{1.00\textwidth}
        \centering
        \includegraphics[width=\linewidth]{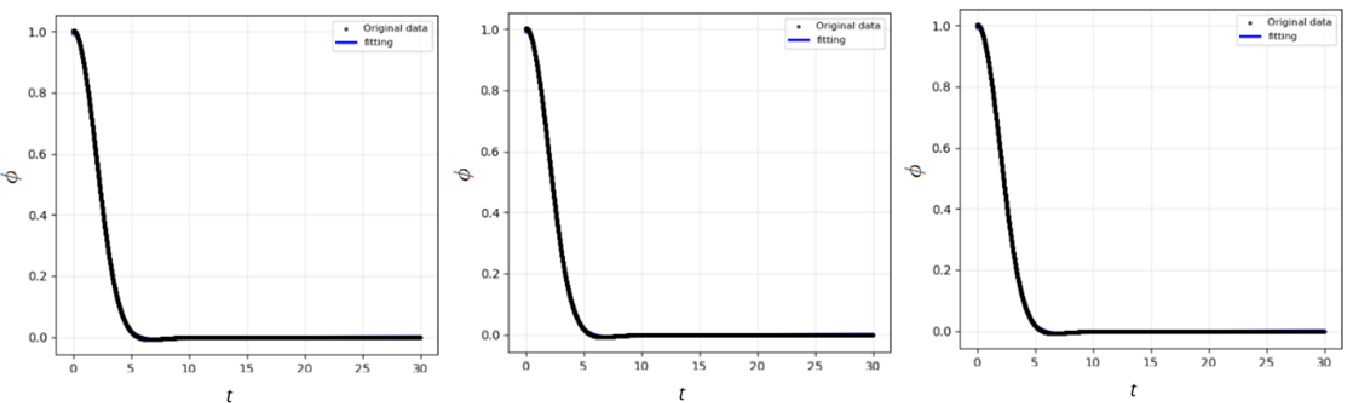}
        \caption{Fit for the scalar field: $\phi(t) = 1.7109t^{4}+2.4664t^{3}+2.2107t^{2}-0.0355t+1.5764$, with statistical metric of $R^2 = 0.99986228$.}
          \label{figura_14}
    \end{minipage}
  \end{figure}

\begin{figure}[!htbp]
    \centering
    \begin{minipage}{0.45\textwidth}
        \centering
        \includegraphics[width=\linewidth]{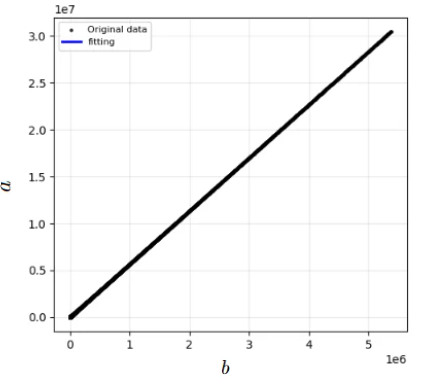}
    \end{minipage}
    \caption{Graph of the best curve, among three obtained via fitting, for the original data (in black) and the fit (blue). The best equation is $a(b) = 5.6722 b - 0.7489$, with statistical metric of $R^2 = 1.00000000$.}
    \label{figura_15}
\end{figure}

\begin{figure}[H]
    \centering
    \begin{minipage}{1.00\textwidth}
        \centering
        \includegraphics[width=\linewidth]{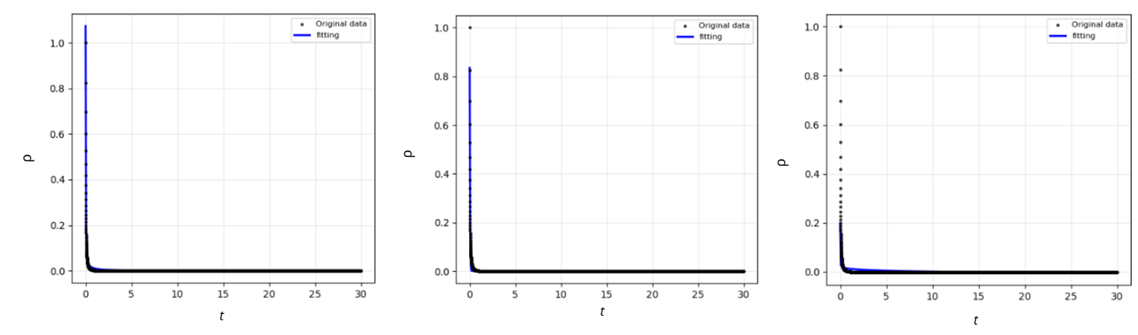}
        \caption{Fit for the density: $\rho(t) = 2545.1909 - 2545.1905\tanh(15.2170 t + 4.3579)$, with statistical metric of $R^2 =  0.96198862$.}
         \label{figura_16}
    \end{minipage}
   
\end{figure}

Here, we used Eqs. (\ref{LRSBIII1}, \ref{LRSBIII2}, \ref{LRSBIII3}) to describe the fitting problem. The corresponding graphs are presented in Figures (\ref{figura_12}), (\ref{figura_13}), (\ref{figura_14}), (\ref{figura_15}) and (\ref{figura_16}), for $a \times t$, $b \times t$, $\phi \times t$, $a \times b$ and $\rho \times t$, respectively.

One of the interesting points to discuss are the $a \times b$ curves found (\ref{figura_4}), (\ref{figura_9}), and (\ref{figura_15}), which describe different graphs and functions, when restored by the fitting, of the \textit{ansatz} that we found. The first case, KS, is a fourth-degree function, but with very small coefficients, making the first-degree polynomial term evident. Similarly, this occurs for the second case, Bianchi I. For Bianchi III, the problem gave a fitting again the same word of a first-degree polynomial, with a metric of 100\% , that is, a straight line.

%%%%%%%%%%%%%%%%%%%%%%%%%%%%%%%%%%%%%%%%%%%%%%%%%%%%%%%%%%%%%%%%%%%%%%%%
\section{Dynamical Aspects}
%%%%%%%%%%%%%%%%%%%%%%%%%%%%%%%%%%%%%%%%%%%%%%%%%%%%%%%%%%%%%%%%%%%%%%%%

In this section, we will verify the dynamical parameters that describe the evolution of the model, as they are important measures for the validation process provided by the Fitting.
\subsection*{Hubble and Shear Parameters}

From the covariant derivative of the four-velocity of the fluid constituting the model, we can determine some kinematic variables that help us understand the development of the model and its connection with the studied universe. We have, then,
\begin{equation}
U_{\alpha ; \beta} = \dot{U}_{\alpha}U_{\beta} + \sigma_{\alpha \beta} + \omega_{\alpha \beta} + \frac{1}{3} \Theta(g_{\alpha \beta} - U_{\alpha}U_{\beta})\quad,
\end{equation}
where $\dot{U}_{\alpha}$ is the acceleration vector, $\sigma_{\alpha \beta}$ is the shear tensor, $\omega_{\alpha \beta}$ is the vorticity tensor (or \textit{twist}), and $\Theta$ is the scalar expansion. For the metrics we are analyzing, we have $\dot{U}_{\alpha} = 0 = \omega_{\alpha \beta}$. The other quantities do not change their forms for the three metrics. Thus, the scalar expansion is given by
\begin{eqnarray}
\Theta = {U^{\alpha}}_{; \alpha} = \frac{\dot{a}}{a} + 2\frac{\dot{b}}{b}\quad. \label{theta}
\end{eqnarray}
The shear tensor, defined as
\begin{equation}
\sigma_{\alpha \beta} = U_{(\alpha ; \beta)} - \dot{U}_{(\alpha}U_{\beta)} - \frac{1}{3} \Theta(g_{\alpha \beta} - U_{\alpha}U_{\beta})\quad,
\end{equation}
provides the shear scalar equal to
\begin{eqnarray}
\sigma = \left[\frac{1}{2} \sigma_{\alpha \beta} \sigma^{\alpha \beta} \right]^{1/2} = \frac{1}{\sqrt{3}} \left|\frac{\dot{a}}{a} - \frac{\dot{b}}{b}\right|\quad. \label{sigma}
\end{eqnarray}

\begin{figure}[!htbp]
\centering
\begin{minipage}{0.48\textwidth}
\centering
\includegraphics[width=\textwidth]{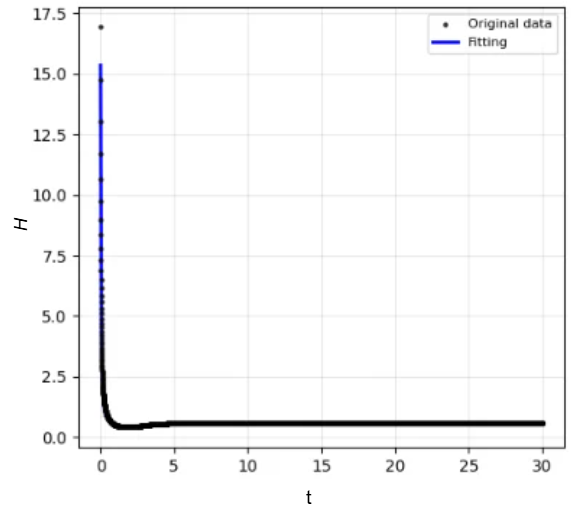} \\[2mm]
\textbf{(a)}
\end{minipage}
\hfill
\begin{minipage}{0.49\textwidth}
\centering
\includegraphics[width=\textwidth]{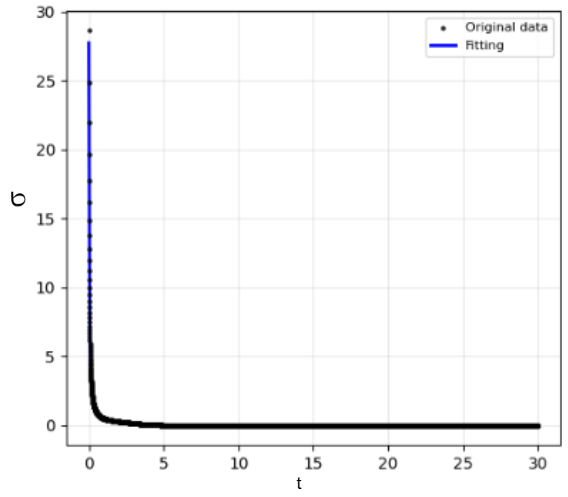}
\textbf{(b)}\\[2mm]
\end{minipage}
    \caption{Comparison between the curves obtained via fitting: original data (in black) and fit (blue) for $H$ (left) and $\sigma$ (right), for KS, with $w = 1/3$.
    (a)   Hubble parameter $H \times t$.
    (b) Shear scalar $\sigma \times t$.}

    \label{figura_17}
\end{figure}

\begin{figure}[!htbp]
    \centering
    \begin{minipage}{0.47\textwidth}
        \centering
       
        \includegraphics[width=\textwidth]{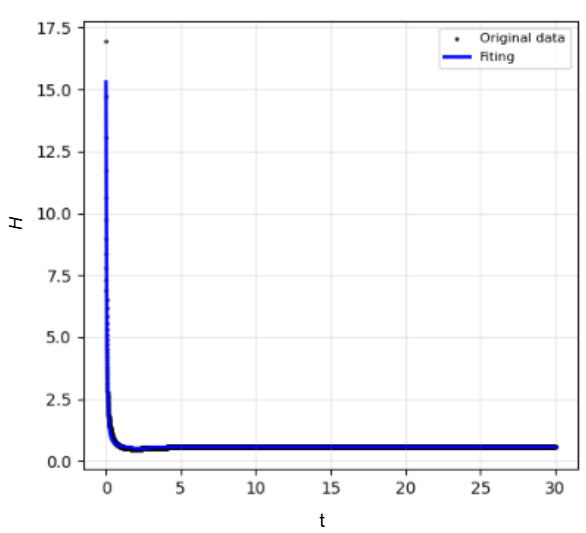} \\[2mm]
         \textbf{(a)}
    \end{minipage}
    \hfill
    \begin{minipage}{0.50\textwidth}
        \centering
         \includegraphics[width=\textwidth]{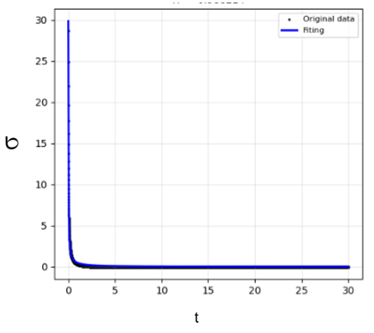}
        \textbf{(b)}\\[2mm]
    \end{minipage}
    \caption{Comparison between the curves obtained via fitting: original data (in black) and fit (blue) for $H$ (left) and $\sigma$ (right), for Bianchi I, with $w = 1/3$.
    (a)   Hubble parameter $H \times t$.
    (b) Shear scalar $\sigma \times t$.}

    \label{figura_18}
\end{figure}

\begin{figure}[h!]
    \centering
    \begin{minipage}{0.48\textwidth}
        \centering       
        \includegraphics[width=\textwidth]{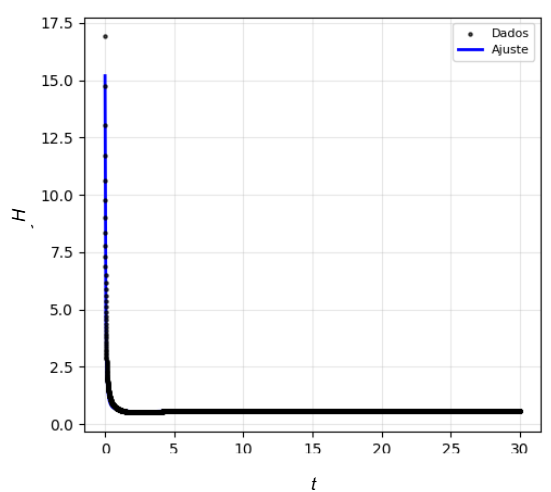} \\[2mm]
         \textbf{(a)}
    \end{minipage}
    \hfill
    \begin{minipage}{0.50\textwidth}
        \centering
         \includegraphics[width=\textwidth]{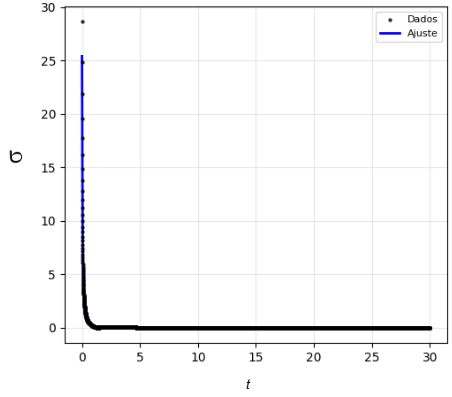}
        \textbf{(b)}\\[2mm]
    \end{minipage}
    \caption{Comparison between the curves obtained via fitting: original data (in black) and fit (blue) for $H$ (left) and $\sigma$ (right), for Bianchi III, with $w = 1/3$.
    (a)   Hubble parameter $H \times t$.
    (b) Shear scalar $\sigma \times t$.}

    \label{figura_19}
\end{figure}

In order to verify the quantities described in this section with the fitting analysis, from the fundamental parameters obtained, the comparative plots of $H \times t$ and $\sigma \times t$ can be verified: for KS in Figure (\ref{figura_17}), for Bianchi I in (\ref{figura_18}), and for Bianchi III in (\ref{figura_19}).
%%%%%%%%%%%%%%%%%%%%%%%%%%%%%%%
%%%%%%%%%%%%%%%%%%%%%%%%%%%%%%%
%%%%%%%%%%%%%%%%%%%%%%%%%%%%%%%%

\subsection*{Density Parameters}

The equality of the expressions for the three metrics stems from the fact that the difference between them lies only in the curvature term, thus maintaining the expressions related to expansion and anisotropy. This can be verified as follows: taking the field equations derived from the Kantowski-Sachs metric, equations \eqref{alpharho1} and \eqref{alpharho2}, from the LRS Bianchi I metric, equations \eqref{LRSBI1} and \eqref{LRSBI2}, and from the LRS Bianchi III metric, equations \eqref{LRSBIII1} and \eqref{LRSBIII2}, we can combine them into two general expressions:
\begin{eqnarray}
2 \frac{\dot{a} \dot{b}}{a b} + \left( \frac{\dot{b}}{b} \right)^2 + \frac{\kappa}{b^2} &=& \rho + \rho_{\phi}\quad, \label{eq1} \\
2\frac{\ddot{b}}{b} + \left(\frac{\dot{b}}{b} \right)^2 + \frac{\kappa}{b^2} &=& -p - p_{\phi}\quad, \label{eq2}
\end{eqnarray}
where $\kappa = 1$ represents a universe with closed geometry (Kantowski-Sachs), $\kappa = 0$ represents a universe with flat geometry (LRS Bianchi I), and $\kappa = -1$ represents a universe with open geometry (LRS Bianchi III).

Rewriting expression \eqref{eq1} in terms of kinematic quantities, equations \eqref{theta} and \eqref{sigma}, we obtain
\begin{equation}
\frac{{\Theta}^2}{3} - {\sigma}^2 + \frac{\kappa}{b^2} = \rho + \rho_{\phi}\quad.
\end{equation}
Knowing the relation between the Hubble parameter and the scalar expansion, $H = \Theta/3$, we can define the density parameters from the above equation, which becomes
\begin{equation}
\Omega_{tot} = \Omega_m + {\Omega}_{\sigma} + {\Omega}_{\kappa} + {\Omega}_{\phi} = 1\quad,
\end{equation}
where
\begin{eqnarray}
\Omega_m = \frac{\rho}{3H^2}, \qquad {\Omega}_{\sigma} = \frac{{\sigma}^2}{3H^2}, \qquad
{\Omega}_{\kappa} = -\frac{\kappa/b^2}{3H^2}, \qquad {\Omega}_{\phi} = \frac{\rho_{\phi}}{3H^2}.
\end{eqnarray}
These parameters help us study the structures of our model, allowing us to understand the influences of different elements in the total composition.

\begin{figure}[!htbp]
    \centering
    \begin{minipage}{0.48\textwidth}
        \centering       
        \includegraphics[width=\textwidth]{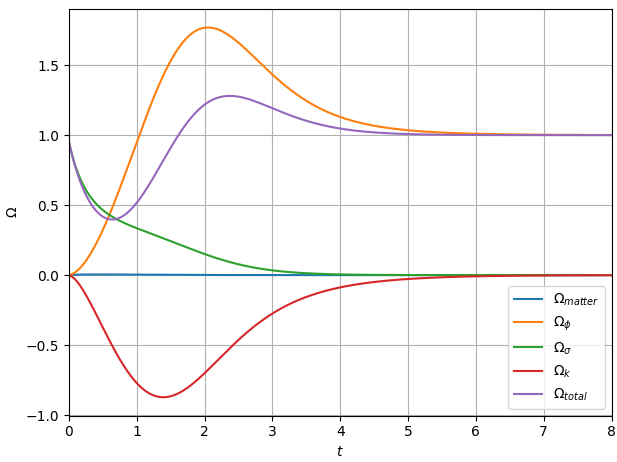} \\[2mm]
         \textbf{(a)}
    \end{minipage}
    \hfill
    \begin{minipage}{0.48\textwidth}
        \centering
         \includegraphics[width=\textwidth]{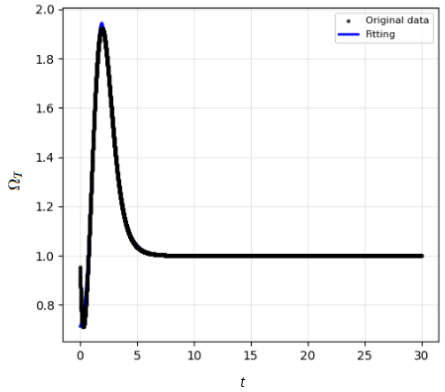}
        \textbf{(b)}\\[2mm]
    \end{minipage}
    \caption{
    (a)   Total density parameter $\Omega_T$ for the KS model. Comparison between the $\Omega$ curves.
    (b) $\Omega_T$ curve of the original data (in black) and the fit (blue), for the KS metric, with $w = 1/3$.}    
    \label{figura_20}
\end{figure}

\begin{figure}[!htbp]
    \centering
    \begin{minipage}{0.48\textwidth}
        \centering       
        \includegraphics[width=\textwidth]{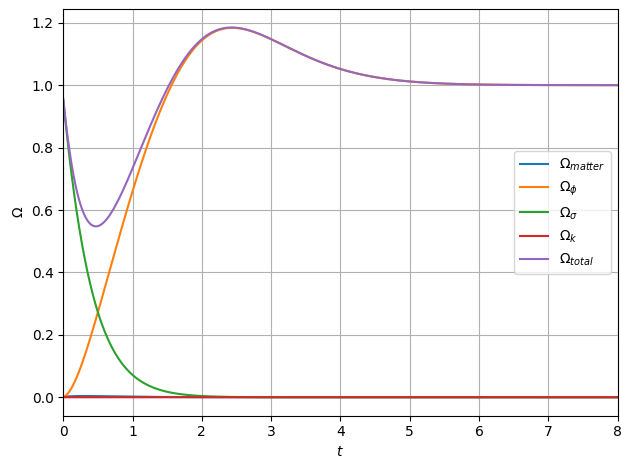} \\[2mm]
         \textbf{(a)}
    \end{minipage}
    \hfill
    \begin{minipage}{0.48\textwidth}
        \centering
         \includegraphics[width=\textwidth]{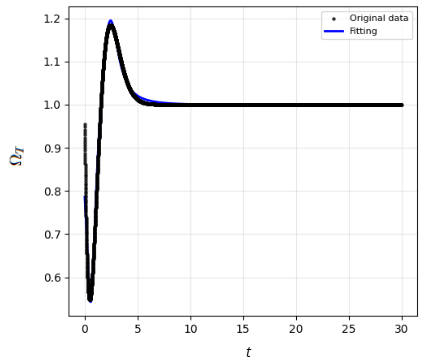}
        \textbf{(b)}\\[2mm]
    \end{minipage}
    \caption{Total density parameter $\Omega_T$ for the Bianchi I model.
    (a)   comparison between the $\Omega$ curves.
F    (b) $\Omega_T$ curve of the original data (in black) and the fit (blue), for the Bianchi I metric, with $w = 1/3$.}    
    \label{figura_21}
\end{figure}

\begin{figure}[!htbp]
    \centering
    \begin{minipage}{0.48\textwidth}
        \centering       
        \includegraphics[width=\textwidth]{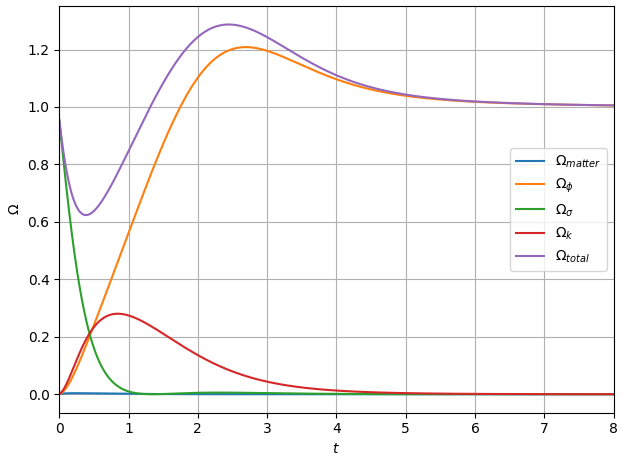} \\[2mm]
         \textbf{(a)}
    \end{minipage}
    \hfill
    \begin{minipage}{0.48\textwidth}
        \centering
         \includegraphics[width=\textwidth]{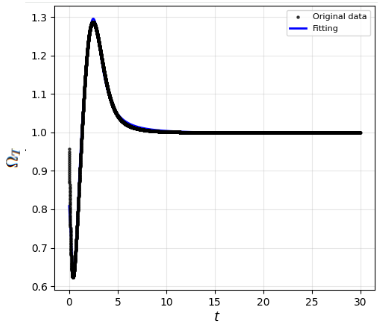}
        \textbf{(b)}\\[2mm]
    \end{minipage}

    \caption{Total density parameter $\Omega_T$ for the Bianchi III model.
    (a)   comparison between the $\Omega$ curves.
    (b) $\Omega_T$ curve of the original data (in black) and the fit (blue), for the Bianchi III metric, with $w = 1/3$.}    
    \label{figura_22}
\end{figure}

In Figures (\ref{figura_20}), (\ref{figura_21}), and (\ref{figura_22}), we present the results for the density parameters for the KS, Bianchi I, and Bianchi III models, respectively, for the case $w = 1/3$.
%%%%%%%%%%%%%%%%%%%%%%%%%%%%%%%%%%%%%%%%%%
%%%%%%%%%%%%%%%%%%%%%%%%%%%%%%%%%%%%%%%%%%
%%%%%%%%%%%%%%%%%%%%%%%%%%%%%%%%%%%%%%%%%
    
%%%%%%%%%%%%%%%%%%%%%%%%%

\subsection*{Expansion Parameters}

Relevant terms for our discussion are those related to the expansion of the studied universe model. The strong energy condition (SEC) is defined, for the three metrics, as
\begin{equation}
R_{\alpha\beta} U^{\alpha} U^{\beta} = -\left(\frac{\ddot{a}}{a} + 2\frac{\ddot{b}}{b} \right)\quad.
\end{equation}
When this quantity assumes negative values, the SEC indicates that there is an accelerated expansion of the universe.

Another parameter that helps us observe whether the model is in accelerated expansion is the deceleration parameter. When negative, it means that the expansion of the universe is accelerated. It is defined as
\begin{equation}
q = -\frac{\ddot{l} \, l}{{\dot{l}}^2}\quad,
\end{equation}
where the average radius is $l = (ab^2)^{1/3}$. Due to all the analyses shown, we see no need to show that the fitting of $q$ was also highly satisfactory.

\begin{figure}[!h]
    \centering
    \begin{minipage}{0.48\textwidth}
        \centering       
        \includegraphics[width=\textwidth]{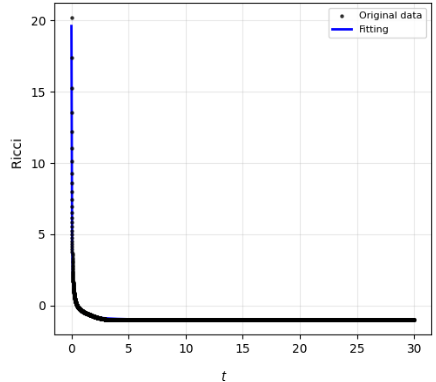} \\[2mm]
         \textbf{(a)}
    \end{minipage}
    \hfill
    \begin{minipage}{0.48\textwidth}
        \centering
         \includegraphics[width=\textwidth]{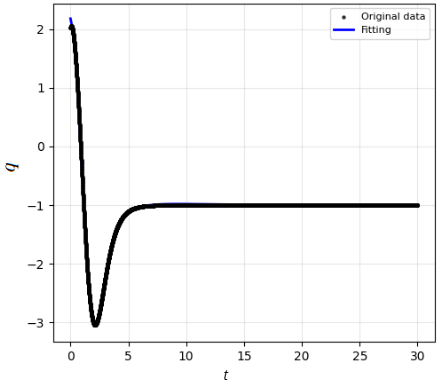}
        \textbf{(b)}\\[2mm]
    \end{minipage}
    \caption{Deceleration parameter $q \times t$ for the KS model.
    (a)   Ricci scalar $R_{\alpha\beta}U^\alpha U^\beta$ as a function of $t$.
    (b) the parameter $q$ for the KS metric, with $w = 1/3$}
    \label{figura_23}
\end{figure}

\begin{figure}[!htbp]
    \centering
    \begin{minipage}{0.48\textwidth}
        \centering       
        \includegraphics[width=\textwidth]{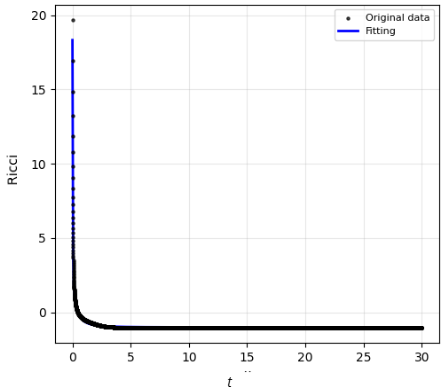} \\[2mm]
         \textbf{(a)}
    \end{minipage}
    \hfill
    \begin{minipage}{0.48\textwidth}
        \centering
         \includegraphics[width=\textwidth]{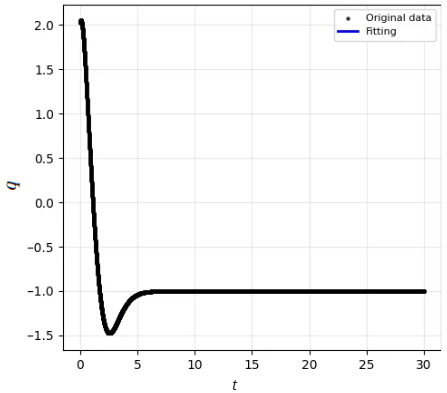}
        \textbf{(b)}\\[2mm]
    \end{minipage}
    \caption{Comparison between the curves for the Bianchi I metric. 
    (a) Ricci scalar $R_{\alpha\beta}U^\alpha U^\beta$ as a function of $t$,
    with the original data in black and the fit in blue;
    (b) deceleration parameter $q$ as a function of $t$, for $w=1/3$.}
    \label{figura_24}
\end{figure}

\begin{figure}[!h]
    \centering
    \begin{minipage}{0.48\textwidth}
        \centering       
        \includegraphics[width=\textwidth]{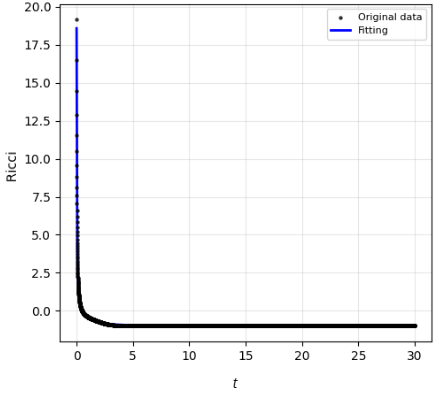} \\[2mm]
         \textbf{(a)}
    \end{minipage}
    \hfill
    \begin{minipage}{0.48\textwidth}
        \centering
         \includegraphics[width=\textwidth]{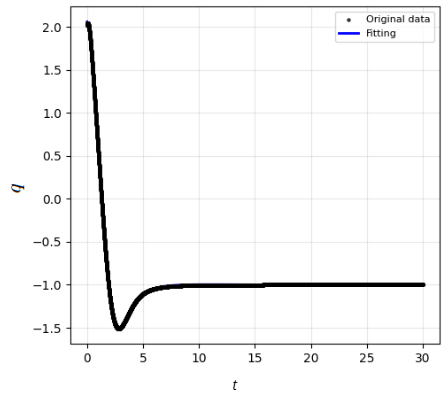}
        \textbf{(b)}\\[2mm]
    \end{minipage}
    \caption{Comparison between the curves for the Bianchi III metric:
    (a) Ricci scalar $R_{\alpha\beta}U^\alpha U^\beta$ as a function of $t$,
    with the original data in black and the fit in blue;
    (b) deceleration parameter $q$ as a function of $t$, for $w=1/3$.}

    \label{figura_25}
\end{figure}

In Figures (\ref {figura_23}), (\ref{figura_24}), and (\ref{figura_25}), we present the results for the Ricci scalar and the deceleration parameter for the KS, Bianchi I, and Bianchi III models, respectively, for the case $w = 1/3$.
%%%%%%%%%%%%%%%%%%%%%%%%%%%%%%%%%%%%%%%%%%%%%%%%%%%%%%%%%%%%%%%%%%%%%%%%%%%%%%%%%%%%%%%%%%%%

%%%%%%%%%%%%%%%%%%%%%%%%%%%%%%%%%%%%%%%%%%%%%%%%%%%%%%%%%%%%%%%%%%%%%%%%%%%%%%%%%%%%%%%%%%%%%%%%%%%%%

\newpage

%%%%%%%%%%%%%%%%%%%%%%%%%%%%%%%%%%%%%%%%%%%%%%%%%%%%%%%%%%%%%%%%%%%%%%%%
\section{Final Considerations}
\label{CC}
%%%%%%%%%%%%%%%%%%%%%%%%%%%%%%%%%%%%%%%%%%%%%%%%%%%%%%%%%%%%%%%%%%%%%%%%

We present, below, the tables with the best fitting functions obtained for the fundamental variables in each of the models and for each type of fluid considered.

\begin{table}[ht]
    \centering
    \caption{Best analytical fitting functions obtained for the Kantowski-Sachs metric.}
    \label{tab:KS}
    \renewcommand{\arraystretch}{0.8}
    \begin{tabular}{c|c|c}
        \toprule
        \textbf{Fluid} & \textbf{Variable} & \textbf{Best fitted analytical expression} \\
        \midrule
        \multirow{4}{*}{$w=\frac13$}
        & $a(t)$ & $0.47102\,\exp(0.57736 t)-1.07901$\\
        & $b(t)$ & $0.33329\,\exp(0.57736\,t)-0.69682$ \\
        & $\phi(t)$ & $1.8155t^3 + 2.3312t^2 + 1.8282t - 0.0303$ \\
        & $\rho(t)$ & $3777.1395 - 3777.1391\,\tanh(15.1537t + 4.5561)$\\
        \midrule
        \multirow{4}{*}{$w=1$}
        & $a(t)$ & $0.4738\,\exp(0.5774t)-0.3293$\\
        & $b(t)$ & $0.3310\,\exp(0.5774t)-0.0688$ \\
        & $\phi(t)$ & $-0.6401t^{3}+2.7443t^{2}+3.4669t-0.3226$ \\
        & $\rho(t)$ & $16.2500t^{2}-0.3897t-0.7645$ \\
        \midrule
        \multirow{4}{*}{$w=0$}
        & $a(t)$ & $0.4640\,\exp(0.5775t)+103.7400$ \\
        & $b(t)$ & $0.3377\,\exp(0.5775t)+75.6360$ \\
        & $\phi(t)$ & $-1.6848\sin(0.3687t-1.3815)-0.1240$ \\
        & $\rho(t)$ & $-14288.9409t^{3}-0.0939t^{2}+0.0117t-2.1632$\\
        \midrule
        \multirow{4}{*}{$w=-1$}
        & $a(t)$ & $0.2457\,\exp(0.8166t)+21944.1162$ \\
        & $b(t)$ & $0.5134\,\exp(0.8166t)+45871.0171$ \\
        & $\phi(t)$ & $1.0186t^{2}+1.4780t+5.4152$ \\
        & $\rho(t)$ & $1.000$ \\
        \bottomrule
    \end{tabular}
\end{table}

\begin{table}[ht]
    \centering
    \caption{Best analytical fitting functions obtained for the Bianchi I LRS metric.}
    \label{tab:BI}
    \renewcommand{\arraystretch}{0.6}
    \begin{tabular}{c|c|c}
        \toprule
        \textbf{Fluid} & \textbf{Variable} & \textbf{Best fitted analytical expression} \\
        \midrule
        \multirow{4}{*}{$w=\frac13$}
        & $a(t)$ & $0.2686\,\exp(0.5774t)-0.7769$ \\
        & $b(t)$ & $0.6778\,\exp(0.5774 t)-1.8972$ \\
        & $\phi(t)$ & $-0.0010+0.9967\,\exp\left[-0.1524( t-0.0871)^2\right]$\\
        & $\rho(t)$ & $ 4348.7578 - 4348.7574\tanh(15.1842 t + 4.6262)$\\
        \midrule
        \multirow{4}{*}{$w=1$}
        & $a(t)$ & $0.2691\,\exp(0.5774 t)-0.2347$ \\
        & $b(t)$ & $0.6767\,\exp(0.5774t)-0.4923$ \\
        & $\phi(t)$ & $-0.0010+0.9968\,\exp\left[-0.1524(t-0.0867)^2\right]$ \\
        & $\rho(t)$ & $4275.6155-4275.6154\,\tanh(31.2973t+4.5724)$ \\
        \midrule
        \multirow{4}{*}{$w=0$}
        & $a(t)$ & $0.2677\,\exp(0.5774t)+19.8316$ \\
        & $b(t)$ & $0.6797\,\exp(0.5774t)+50.3627$ \\
        & $\phi(t)$ & $1.7628\sin(0.5412t-3.4846)-5.3124$ \\
        & $\rho(t)$ & $3671.3508 - 3671.3499\tanh(7.9744 t + 4.5946)$ \\
        \midrule
        \multirow{4}{*}{$w=-1$}
        & $a(t)$ & $0.2006\,\exp(0.8165t)+9623.4314$\\
        & $b(t)$ & $0.6653\,\exp(0.8165t)+31894.5688$\\
        & $\phi(t)$ & $1.1165t^{3}-0.5529t^{2}+2.9259t-0.0311$ \\
        & $\rho(t)$ & $1.000$ \\
        \bottomrule
    \end{tabular}
\end{table}

\begin{table}[ht]
    \centering
    \caption{Best analytical fitting functions obtained for the Bianchi III LRS metric.}
    \label{tab:BIII}
    \renewcommand{\arraystretch}{0.6}
    \begin{tabular}{c|c|c}
        \toprule
        \textbf{Fluid} & \textbf{Variable} & \textbf{Best fitted analytical expression} \\
        \midrule
        \multirow{4}{*}{$w=\frac13$}
        & $a(t)$ & $0.1614\,\exp(0.5774t)-0.7514$\\
        & $b(t)$ & $0.9157\,\exp(0.5774t)-5.0058$ \\
        & $\phi(t)$ & $1.7109t^{4}+2.4664t^{3}+2.2107t^{2}-0.0355t+1.5764$\\
        & $\rho(t)$ & $2545.1909 - 2545.1905\tanh(15.2170 t + 4.3579)$\\
        \midrule
        \multirow{4}{*}{$w=1$}
        & $a(t)$ & $0.1617\,\exp(0.5774t)-0.0268$ \\
        & $b(t)$ & $0.9153\,\exp(0.5774t)-0.5756$ \\
        & $\phi(t)$ & $1.5180t^{3}-0.4637t^{2}+1.5736t-0.0840$ \\
        & $\rho(t)$ & $4030.3870-4030.3869\,\tanh(31.3151t+4.5428)$\\
        \midrule
        \multirow{4}{*}{$w=0$}
        & $a(t)$ & $0.1614\,\exp(0.5774t)+9.1321$ \\
        & $b(t)$ & $0.9178\,\exp(0.5774t)+51.0818$ \\
        & $\phi(t)$ & $1.7968t^{2}+2.4669t+2.2108$ \\
        & $\rho(t)$ & $-11741.9389t^{3}-0.0821t^{2}+0.0104t-0.4734$ \\
        \midrule
        \multirow{4}{*}{$w=-1$}
        & $a(t)$ & $0.1520\,\exp(0.8166t)+15108.4012$\\
        & $b(t)$ & $0.7885\,\exp(0.8166t)+78473.7188$ \\
        & $\phi(t)$ & $-0.3504t^{3}+1.4708t^{2}+6.5134t-0.3985$\\
        & $\rho(t)$ & $1.0000$\\
        \bottomrule
    \end{tabular}
\end{table}

In this work, we used the Einstein equations to describe a cosmological model whose scenario is given by the Kantowski-Sachs metric, that is, a homogeneous and anisotropic universe whose spatial section has the topology $\mathbb{R} \times S^{2}$. We derived the field equation considering the material content of the universe in a comprehensive manner, taking into account the effect of all phases of the universe's evolution, such as radiation, stiff matter, dust, vacuum, etc., without the need to analyze, beforehand, these evolution phases on a case-by-case basis.

Unlike many current works that use dynamical systems to analyze the stability of solutions of anisotropic cosmological models, our development of the calculations presented here demonstrates that the simplicity of the Kantowski-Sachs cosmological model with a general expression for the material content of the universe offers a valuable opportunity to solve and understand analytically the evolution of the basic scenario throughout cosmic history.

It is also important to highlight that, unlike other articles related to this subject discussed here, our calculations and solutions do not use schemes that remove terms from the Einstein equations to avoid divergences or other mathematical irregularities. We refer specifically to the term $1/b^2$. As can be seen in the work presented above, we used all expressions and terms to present new solutions.

Here, we also calculated the cosmological parameters of the model that describe the global dynamics of the Universe, such as its expansion rate and the matter content of some of its constituents. We used the parameters presented in the Raychaudhuri equation to verify the behavior of geodesic lines for more or less anisotropic models. If, however, the observational data suggesting a violation of isotropy on large scales are better observed, the scenario discussed here is a good and viable theoretical model to present as an alternative to the $\Lambda$CDM model \cite{Migkas, costa, land, eriksen, hansen, king, weeb, shamira, shamirb, avelino, mariano, Do}.

Our research assumes significant importance by providing a more precise understanding of cosmic evolution, especially by recognizing and studying anisotropy effects that may be present currently in the universe. By addressing these anisotropies, we contribute to the development of models that fit more of the observations made. Furthermore, by assigning a value to the amount of anisotropy considered, represented by the index $n$, we are able to directly quantify the degree of deviation existing in a specific model, allowing a clear and objective representation of this rate of variation, also providing a link with the currently available observational data.

Finally, future works are in our plans, using this new, more general method and without conditions established a priori. Analyses of perturbations in matter density and the formation of large-scale structures related to dark matter and dark energy, carried out together with observations of the cosmic microwave background in the context of anisotropic models, quite developed and precise, will be redone with the theoretical tools developed here in this study. We also intend to develop works involving tensor perturbations, the cosmological origin, and the temporal evolution of gravitational waves in models that deviate from the traditional paradigm of isotropy, with the knowledge acquired here without impositions that limit the analysis. We also intend to deepen studies involving the variations found in observations of large-scale isotropy of the universe and the relation with the index $n$ of our cosmological model.

%%%%%%%%%%%%%%%%%%%%%%%%%%%%%%%%%%%%%%%%%%%%%%%%%%%%%%%%%%%%%%%%%%%%%%%%
\section*{Author Contributions}
%%%%%%%%%%%%%%%%%%%%%%%%%%%%%%%%%%%%%%%%%%%%%%%%%%%%%%%%%%%%%%%%%%%%%%%%

All authors contributed to all stages of the article, including writing.

%%%%%%%%%%%%%%%%%%%%%%%%%%%%%%%%%%%%%%%%%%%%%%%%%%%%%%%%%%%%%%%%%%%%%%%%
\section*{Acknowledgments}
%%%%%%%%%%%%%%%%%%%%%%%%%%%%%%%%%%%%%%%%%%%%%%%%%%%%%%%%%%%%%%%%%%%%%%%%

We are particularly grateful to Germano A. Monerat for his suggestions and valuable discussions. The authors also thank the computational resources of LFTC-IFES, Campus Cariacica and Sci-Com Lab/DFis/UFES.

%%%%%%%%%%%%%%%%%%%%%%%%%%%%%%%%%%%%%%%%%%%%%%%%%%%%%%%%%%%%%%%%%%%%%%%%
\section*{Declarations}
%%%%%%%%%%%%%%%%%%%%%%%%%%%%%%%%%%%%%%%%%%%%%%%%%%%%%%%%%%%%%%%%%%%%%%%%

The authors declare that:
\begin{enumerate}
    \item \textbf{Ethical Approval}: not applicable to our research presented here.
    \item \textbf{Conflicts of Interest}: not applicable to our research presented here.
    \item \textbf{Funding}: not applicable to our research presented here.
    \item \textbf{Data and Materials Availability}: not applicable to our research presented here.
\end{enumerate}

%%%%%%%%%%%%%%%%%%%%%%%%%%%%%%%%%%%%%%%%%%%%%%%%%%%%%%%%%%%%%%%%%%%%%%%%

\end{document}